\documentclass[11pt]{article}
\usepackage[utf8]{inputenc}  
\usepackage[english]{babel}    
\usepackage{hyperref}        
\usepackage{booktabs}		 
\usepackage{a4wide}          
\usepackage{amsmath,amsfonts}
\usepackage{graphicx}
\usepackage{xcolor}
\usepackage{algorithmic}
\usepackage{algorithm}

\title{Neural Network–Assisted RIS Weight Optimization for Spatial Nulling in Distorted Reflector Antenna Systems}
\author{XINRUI LI, R. MICHAEL BUEHRER}

\begin{document}

\maketitle                  
\thispagestyle{empty}       


\section*{Abstract}
Reconfigurable intelligent surfaces (RIS) have recently been proposed as an effective means for spatial interference suppression in large reflector antenna systems. Existing RIS weight optimization algorithms typically rely on accurate theoretical radiation models. However, in practice, distortions on the reflector antenna may cause mismatches between the theoretical and true antenna patterns, leading to degraded interference cancellation performance when these weights are directly applied. In this report, a residual learning network–assisted simulated annealing (ResNet-SA) framework is proposed to address this mismatch without requiring explicit knowledge of the distorted electric field. By learning the residual difference between the theoretical and true antenna gains, a neural network (NN) is embedded in a heuristic optimization algorithm to find the optimal weight vector. Simulation results demonstrate that the proposed approach achieves improved null depth in the true radiation pattern as compared with conventional methods that optimize weights based solely on the theoretical model, validating the effectiveness of the ResNet-SA algorithm for reflector antenna systems with approximate knowledge of the pattern.

\section{Introduction}
Reconfigurable intelligent surfaces (RIS) have recently emerged as a promising technology for enhancing the performance of wireless communication systems. An RIS typically consists of a surface array of sub-wavelength reflecting elements whose electromagnetic responses can be dynamically adjusted. By appropriately tuning the phase and amplitude of these elements, the RIS can reshape the incident wavefronts and effectively manipulate the wireless propagation environment between transceivers \cite{8796365, 8741198}. \par
Inspired by these advances, RIS can also serve as an effective auxiliary mechanism for spatial interference suppression in large ground-based antenna systems, which commonly employ reflector antennas. Such antennas produce highly directive main lobes toward the desired source, but also generate sidelobes that may capture unwanted interference from other directions. To address these limitations, the authors in \cite{9400735} introduced an approach that integrates the RIS with the reflector edge treatments, where reconfigurable elements are mounted along the rim of a reflector antenna to impose controllable phase shifts on the reflected fields. The motivations for this architecture, along with its advantages and limitations relative to conventional interference mitigation methods, are discussed in Section I of \cite{9400735}. This architecture enables existing radio telescopes to be retrofitted with minimal modifications to the original reflector geometry and mechanical structure. Subsequently, improved algorithms for determining the optimal RIS weights were proposed in \cite{10773626}. Furthermore, the authors in \cite{10403767} presented a practical design for such a reconfigurable antenna and demonstrated the efficacy of this concept through full-wave simulation. \par
Despite these advances, the weight-optimization algorithms in \cite{9400735} and \cite{10773626} rely on the assumption that the true radiation pattern of the reflector antenna exactly matches the theoretical pattern obtained via physical optics. In practice, however, the true electric field distribution is influenced by systematic errors \cite{balanis2016antenna}, which arise from factors such as gravitational deformation, wind loading, temperature variations, ice or snow accumulation, and manufacturing imperfections. Although several studies have addressed reflector distortion modeling \cite{5175499, 931261, 7490430}, estimating the distortion parameters and recomputing the RIS weights based on the resulting true electric field intensity substantially increases the computational complexity of the system. On the other hand, although the heuristic algorithm in \cite{10773626} requires only measured antenna gains, it remains impractical because executing the algorithm would require obtaining the gain for every possible weight vector, which must be measured individually in practice and is therefore infeasible due to the large number of weight combinations. \par
Nevertheless, it is important to note that the true radiation pattern typically deviates only slightly from the known theoretical pattern. Moreover, the heuristic optimization framework in \cite{10773626} operates solely on the resulting antenna gain associated with a given weight vector. Therefore, the key unknown quantity is the residual difference between the theoretical and true antenna gains. Residual learning neural networks (ResNets), which are well suited for learning such discrepancies relative to a known reference \cite{he2016deep}, naturally fit this problem. In this work, the ResNet takes the RIS weight vector and the corresponding theoretical antenna gain as inputs and predicts the true antenna gain. This prediction can be incorporated into the heuristic algorithm in \cite{10773626}, which can be performed offline once the NN is trained. In addition, the ResNet can achieve satisfactory performance with a relatively small number of training samples. \par
Based on these observations, we propose a residual learning network–assisted simulated annealing (ResNet-SA) framework as an extension of the algorithms proposed in \cite{10773626} for reflector antennas subject to distortion. Simulation results demonstrate the effectiveness of the proposed approach and highlight its potential for the RIS weight optimization under more general and realistic scenarios.  \par

\section{System Model}
\label{sec:II}
\begin{figure}[t!]
\begin{center}
\includegraphics[scale=0.5]{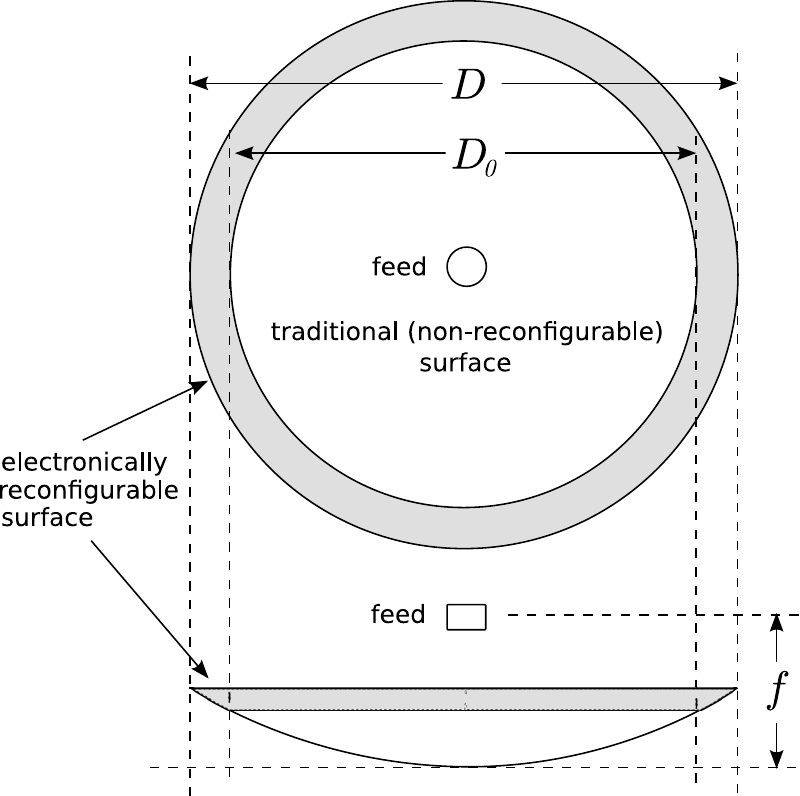}
\includegraphics[scale=0.5]{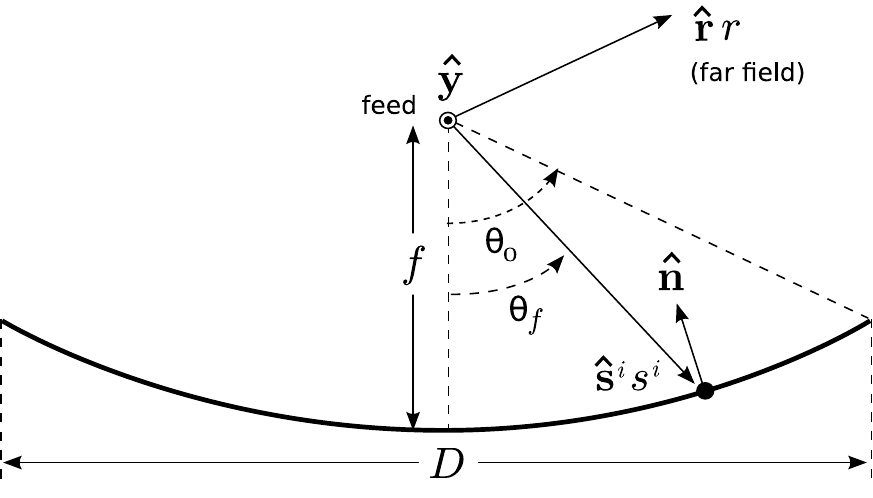}
\caption{On-axis (left) view of an electronically- reconfigurable rim scattering system assumed in this paper (reproduced from \cite{9400735}). The global origin is defined as the bottom of the dish. \(\theta_f\) is the angle measured from the reflector axis of rotation toward the rim with \(\theta_f = \theta_1\) at the rim of the fixed portion of the dish and \(\theta_f = \theta_0\) at the rim of the entire dish, \(\phi\) is the angular coordinate orthogonal to both \(\theta_f\) and the reflector axis. }
\label{fig:system_setting}
\end{center}
\end{figure}

We consider a prime focus axisymmetric circular paraboloidal reflector antenna system equipped with RIS elements along the outer rim, as illustrated in Fig.~\ref{fig:system_setting}. Following the development in~\cite{9400735}, we assume reciprocity between the transmit and receive patterns, and calculate the transmit patterns using physical optics (PO) \cite{stutzman2012antenna}. The total electric field intensity \(\mathbf{E}^s\) scattered by the reflector in the far-field direction with zenith angle \(\psi\) and azimuth angle \(\phi\) is given by 
\begin{equation}
\mathbf{E}^s(\psi, \phi) = \mathbf{E}_f^s(\psi, \phi) + \mathbf{E}_r^s(\psi, \phi),
\end{equation}
where \(\mathbf{E}_f^s\) represents the electric field intensity contributed from the fixed (non-reconfigurable) portion of the reflector (including any negligible feed-direct contribution), and \(\mathbf{E}_r^s\) accounts for the scattered field from the reconfigurable portion of the reflector. Specifically, \(\mathbf{E}_f^s\) can be written as
\begin{equation}
\mathbf{E}_f^s(\psi, \phi) = -j \omega \mu_0 \frac{e^{-j \beta r}}{4 \pi r} \int_{\theta_f = 0}^{\theta_1} \int_{\phi' = 0}^{2 \pi} \mathbf{J}_0 \left(\mathbf{s}^i \right) e^{j \beta \hat{\mathbf{r}} (\psi, \phi) \cdot \mathbf{r}'} \ \mathrm{d}s,
\label{eq:E_f_continue}
\end{equation}
where \(\omega\) is the angular frequency, \(\mu_0\) is the permeability of free space, \(\beta\) is the wavenumber, \(r\) denotes the distance between the origin and the far field observation point in the direction \(\psi\), and \(\mathrm{d}s\) is the differential element of surface area. The PO equivalent surface current distribution for the fixed portion of the dish \(\mathbf{J}_0 (\mathbf{s}^i) = 2 \hat{\mathbf{n}} \times \mathbf{H}^i(\mathbf{s}^i)\) is computed from the incident magnetic field \(\mathbf{H}^i(\mathbf{s}^i)\) and the surface normal \(\hat{\mathbf{n}}\). As in \cite{9400735}, we model the feed as
\begin{equation}
\mathbf{H}^i(\mathbf{s}^i) = I_0 \frac{\hat{\mathbf{y}} \times \hat{\mathbf{s}}^i}{\left| \hat{\mathbf{y}} \times \hat{\mathbf{s}}^i \right|} \frac{e^{-j \beta s^i}}{s^i} \cos(\theta_f)^q, 
\label{eq:mag_field}
\end{equation}
where \(I_0\) denotes the feed magnitude and phase, \(q\) controls the feed directivity, and \(\mathbf{s}^i = \hat{\mathbf{s}}^i s^i\) is the source-to-surface position vector with magnitude \(s^i\) pointing toward the direction \(\hat{\mathbf{s}}^i\). The far field unit vector \(\hat{\mathbf{r}}(\psi, \phi)\) pointing towards the direction \((\psi, \phi)\) is 
\begin{equation}
\hat{\mathbf{r}}(\psi, \phi) = \sin(\psi)\cos(\phi) \ \hat{\mathbf{x}} + \sin(\psi)\sin(\phi) \ \hat{\mathbf{y}} +  \cos(\psi) \ \hat{\mathbf{z}}. 
\end{equation}
These geometric parameters are illustrated in Fig.~\ref{fig:system_setting}. \par
In contrast to the fixed portion of the dish, the electric field intensity scattered by the reconfigurable portion of the reflector is similarly written as
\begin{equation}
\mathbf{E}_r^s(\psi, \phi) = -j \omega \mu_0 \frac{e^{-j \beta r}}{4 \pi r} \int_{\theta_f = \theta_1}^{\theta_0} \int_{\phi' = 0}^{2 \pi} \mathbf{J}_1 \left(\mathbf{s}^i \right) e^{j \beta \hat{\mathbf{r}} (\psi, \phi) \cdot \mathbf{r}'} \ \mathrm{d}s,
\end{equation}
where the major differences between the contributions due to the fixed and reconfigurable portions of the dish are (a) the angles in the integration and (b) the PO equivalent surface current distribution. Due to the discrete nature of the reconfigurable surface, we can write \(\mathbf{E}_r^s(\psi, \phi)\) as a summation over the contributions of individual elements, that is
\begin{equation}
\mathbf{E}_r^s (\psi, \phi) = -j \omega \mu_0 \frac{e^{-j \beta r}}{4 \pi r} \sum_n \mathbf{J}_1 \left(\mathbf{s}_n^i \right) e^{j \beta \hat{\mathbf{r}}(\psi, \phi) \cdot \mathbf{r}_n'} \Delta s,
\label{eq:E_discrete}
\end{equation}
where \(\mathbf{r}_n'\) can be parameterized by the radial distance \(\rho\) instead of the angle \(\theta_f\), and \(\Delta s\) denotes the area associated with each RIS element. \(\mathbf{J}_1 \left(\mathbf{s}_n^i \right) = w_n \mathbf{J}_0 \left(\mathbf{s}_n^i \right)\) is the current distribution due to the \(n\)-th element with complex-valued weight \(w_n\). These weights will be designed to cancel sidelobes in specific directions of the co-polarized (co-pol) pattern. Specifically, we define the complex scalar \(E_f^{s, co} (\psi, \phi)\) and \(E_r^{s, co} (\psi, \phi)\) to represent the magnitude of \(\mathbf{E}_f^s (\psi, \phi)\) and \(\mathbf{E}_r^s (\psi, \phi)\) along the co-pol direction, respectively. The contribution of the \(N\) reconfigurable elements can be compactly written in the form:
\begin{equation}
E_r^{s, co} (\psi, \phi) = \mathbf{e}_{\psi, \phi}^T \mathbf{w},
\end{equation}
where \(\mathbf{w} \in \mathbb{C}^{N}\) denotes the complex RIS weights, and \(\mathbf{e}_{\psi, \phi} \in \mathbb{C}^N\) represents the co-pol portion of the electric field intensity without the influence of the RIS elements. The \(n\)-th element of \(\mathbf{e}_{\psi, \phi}\) is given by
\begin{equation}
e_{\psi, \phi, n} = \left(-j \omega \mu_0 \frac{e^{-j\beta r}}{4 \pi r} \mathbf{J}_0(\mathbf{s}_n^i) e^{j \beta \hat{\mathbf{r}}(\psi, \phi) \cdot \mathbf{r}_n'} \Delta s \right) \cdot \hat{\mathbf{e}}_{\mathrm{co}},
\label{eq:e_discrete}
\end{equation}
where \(\hat{\mathbf{e}}_{\mathrm{co}}\) is the co-polorized direction.
 
\begin{figure}[t!]
\begin{center}
\includegraphics[scale=0.5]{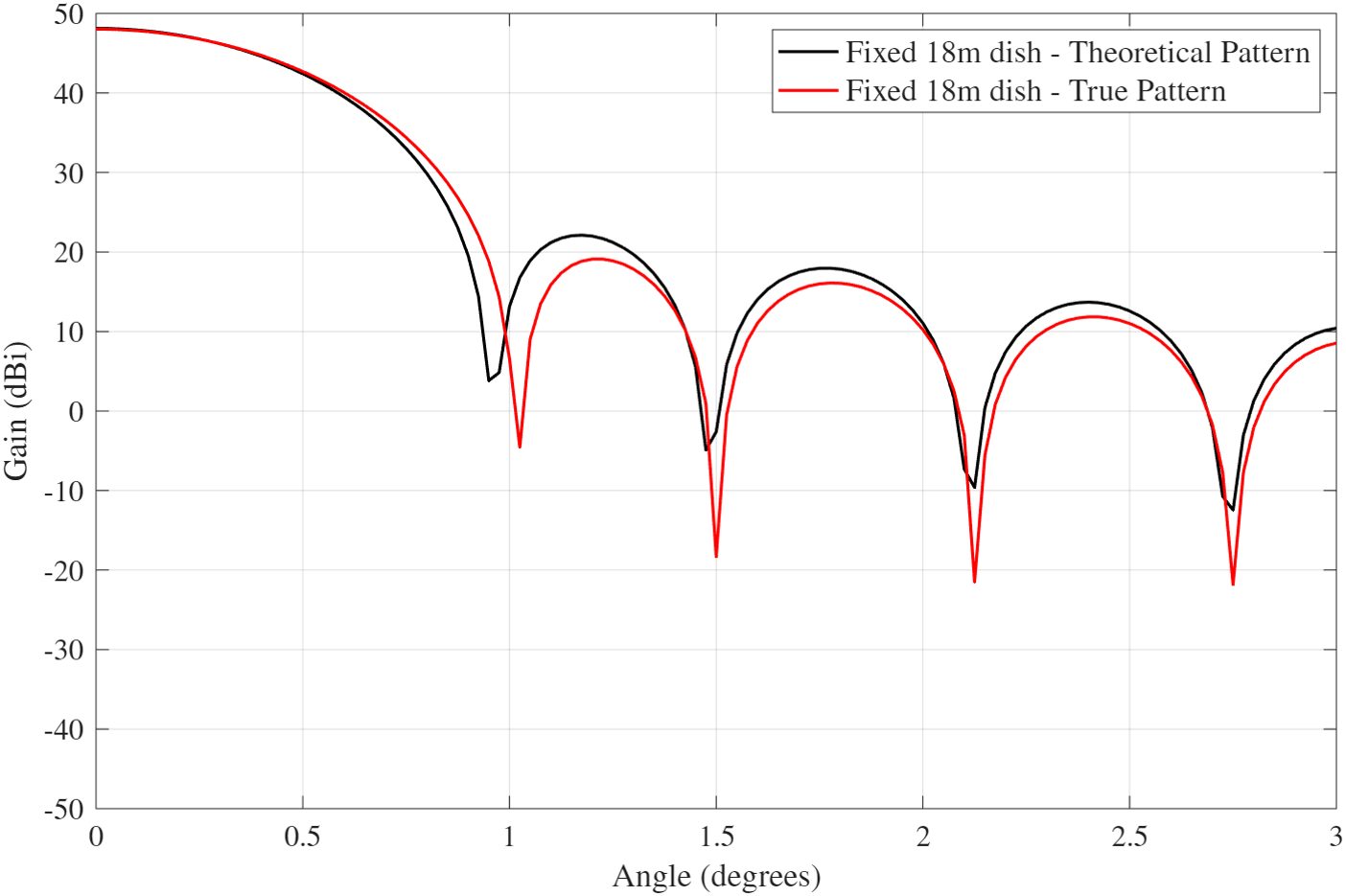}
\caption{Theoretical radiation pattern versus the true radiation pattern on H-plane \(\phi = 0^{\circ}\) with zenith angle ranging from \(\psi = 0^{\circ}\) to \(3^{\circ}\).  }
\label{fig:ideal_vs_true_pattern}
\end{center}
\end{figure}

Now to suppress the sidelobe gain at a specific direction, we desire that
\begin{equation}
E_f^{s, co}(\psi, \phi) + E_r^{s, co}(\psi, \phi) = 0,
\end{equation}
which implies that the electric field intensity from the reconfigurable portion of the reflector must cancel that from the fixed portion of the reflector. Recalling that \(E_r^{s, co} (\psi, \phi) = \mathbf{e}_{\psi, \phi}^T \mathbf{w}\), the goal becomes finding the optimal weight vector \(\mathbf{w}^*\) that satisfies 
\begin{equation}
\mathbf{e}_{\psi, \phi}^T \mathbf{w}^* = -E_f^{s, co}(\psi, \phi).
\label{eq:p_single}
\end{equation}
Due to the presence of systematic distortions, the true electric field intensity of the antenna generally deviates from the theoretical expressions introduced above, resulting in a modified radiation pattern. In this work, the true antenna pattern is generated by perturbing the parameter \(q\) in \eqref{eq:mag_field}. While the true electric field in practice is influenced by other multiple factors, the adopted parametric perturbation provides a convenient and repeatable means of evaluating the proposed algorithm under model mismatch. Importantly, the algorithm itself does not rely on the specific form of this perturbation and can be readily extended to more physically realistic distortion models. Fig.~\ref{fig:ideal_vs_true_pattern} presents a comparison between the theoretical pattern \(q =1.14\) and the distorted pattern \(q = 1.5\) on the H-plane (\(\phi = 0^{\circ}\)) assuming all RIS weights are set to unity. We can see that the resulting radiation patterns exhibit small but noticeable differences for the two selected \(q\)-values. While the differences may seem to be insignificant, they are actually very significant when attempting to create deep nulls in the pattern. 

\section{Weight Selection}
The approach to find the appropriate set of weights which satisfies (\ref{eq:p_single}) depends on the restrictions placed on \(\mathbf{w}^*\). To be more practical, we consider the case when each element is typically constrained to a finite set of discrete phase shifts due to hardware limitations. This motivates of solving the problem of 
\begin{equation}
\begin{split}
\min_{\mathbf{w}\in \mathbb{C}^{N}} \quad & f(\mathbf{w}) = \Vert \mathbf{Aw} - \mathbf{y}\Vert^2 \\
\text{s.t. } \; \quad  &w_n \in \mathcal{W}, \quad n = 1, \dots, N, 
\label{eq:LS2}
\end{split}
\end{equation}
where \(\mathcal{W} = \left\{w \in \mathbb{C} \big| w = e^{j \frac{2 \pi k}{M}}, k = 0, \dots, M-1 \right\}\) and \(M\) is the number of quantization levels. For the theoretical pattern, we addressed this problem using two approaches in our previous work \cite{10773626}. One is using the simulated annealing (SA) algorithm as a representation of the heuristic approaches. Specifically, the phase of one weight element is randomly changed at each step; the algorithm accepts the new weight vector if the cost is lower or accepts the new weight vector with a certain probability if the cost is higher. In \cite{10773626}, we also presented the extreme point pursuit (EXPP) method, which transforms the original problem (\ref{eq:LS2}) into a problem with non-convex objective function but with convex constraint. The new problem is solved using a Majorize-Minimization (MM) algorithm and is guaranteed to converge to a stationary point of the problem (\ref{eq:LS2}). \par
\begin{figure}[t!]
\begin{center}
\includegraphics[scale=0.48]{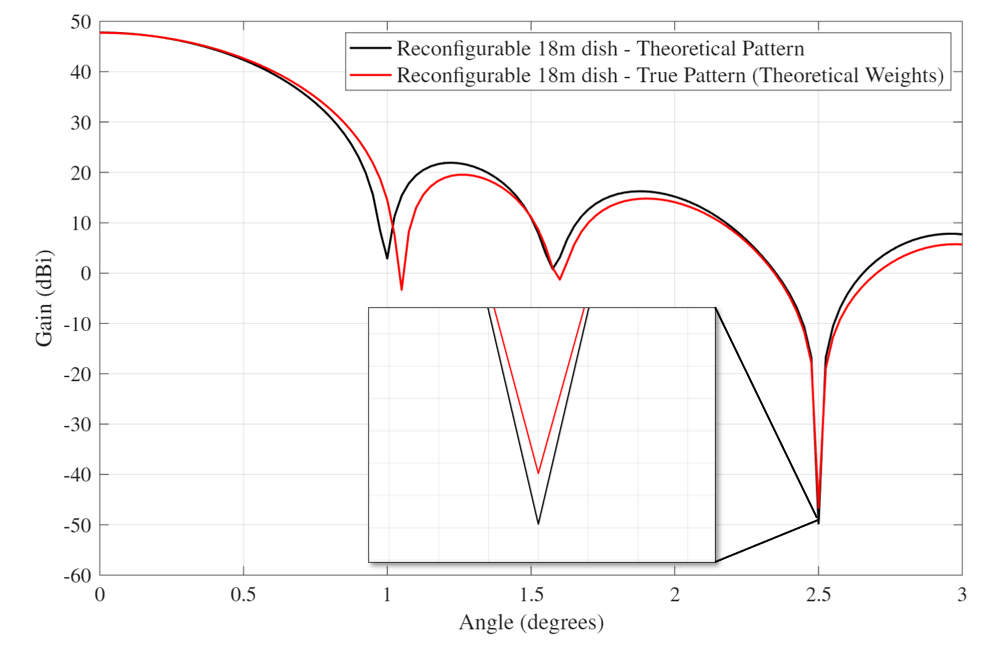}
\caption{Theoretical radiation pattern versus the true radiation pattern on H-plane \(\phi = 0^{\circ}\) with a null placed at \(\psi = 2.5^{\circ}\). The null depth achieved in the true radiation pattern (red curve) is degraded when using the weight vector optimized for the theoretical pattern (black curve). }
\label{fig:pattern_25}
\end{center}
\end{figure}
Although the algorithms presented in \cite{10773626} demonstrate strong performance in identifying RIS weight vectors that yield deep nulls for the theoretical radiation pattern, their effectiveness degrades when applied to the true radiation pattern. This degradation arises from the mismatch between the theoretical and true electric field intensities, which leads to a compromised null depth when the theoretically optimized weights are directly applied. Fig.~\ref{fig:pattern_25} illustrates such a case, where a null is placed at \(\psi = 2.5^{\circ}\). Furthermore, these algorithms are not directly applicable to optimizing the true radiation pattern. In particular, the EXPP algorithm requires explicit expressions of the true electric field intensity vectors to compute the convergence gradient, which are generally unavailable in practice. To address these limitations, we propose an algorithm that leverages NN to determine RIS weight vectors that achieve improved null depth for the true radiation pattern, compared with directly using weights optimized for the theoretical model. \par
Recall that, for the theoretical radiation pattern, the SA algorithm determines whether to accept a candidate weight vector by comparing the resulting antenna gain with that of the current solution. Motivated by this structure, we adopt the SA framework for optimizing the true radiation pattern as well. However, instead of explicitly evaluating the true antenna gain, the gain associated with each candidate weight vector is predicted by a trained NN. Specifically, a residual learning NN is embedded within the SA process to estimate the mismatch between the theoretical and true radiation patterns. \par
We now discuss the principle and the implementation of the ResNet-SA algorithm. During the training phase, we assume that the true radiation pattern can be measured, allowing us to collect a finite set of radiation gain samples at each angle for different RIS weight vectors. This can be done in practice by measuring the received power of the signals from objects at know angles in the pattern. Meanwhile, the corresponding theoretical gains can be computed offline using the known analytical model. Therefore, for each selected weight vector \(\mathbf{w}^{(k)}\), we are able to obtain the theoretical gain \(G^{(k)}\) and the true gain \(\tilde{G}^{(k)}\). The supervised residual learning network is trained to take the weight vector as input and predict the gain mismatch \(\Delta G^{(k)} = \tilde{G}^{(k)} - G^{(k)}\). Once trained, the NN is incorporated into the SA algorithm. The overall SA procedure follows the same structure as in \cite{10773626}, with the key difference that the gain used to decide whether a candidate weight vector is accepted is now obtained from the NN prediction rather than direct evaluation of the true radiation pattern. The complete ResNet-SA framework is summarized in Algorithm~\ref{alg:ResNet_SA}. 
\begin{algorithm}[H]
\caption{Residual Learning Network-Assisted Simulated Annealing (ResNet-SA).}
\begin{algorithmic}
\STATE 
\STATE Initialization: Set \(k\) = 0, \(\mathbf{w}^{(0)}\) randomly chosen from \(\mathcal{W}\). Cooling schedule \(\mathbf{T} = \left[1, \frac{1}{2}, \dots, \frac{1}{T}\right]\)
\STATE \textbf{Repeat: }
\STATE \hspace{0.5cm} \(\mathbf{w}^{(k + 1)} = \mathbf{w}^{(k)}\)
\STATE \hspace{0.5cm} \(n \sim \mathrm{Uniform} \{1, 2, \dots, N \}\), \(m \sim \mathrm{Uniform} \{0, 2, \dots, M-1 \);
\STATE \hspace{0.5cm} \(w_n^{(k + 1)} = e^{j \frac{2\pi m}{M}}\);
\STATE \hspace{0.5cm} Predict \(\Delta \tilde{G}^{(k)}\), \(\Delta \tilde{G}^{(k + 1)}\) using the NN; \(\tilde{G}^{(k)} = G^{(k)} + \Delta \tilde{G}^{(k)}\), \(\tilde{G}^{(k + 1)} = G^{(k + 1)} + \Delta \tilde{G}^{(k + 1)}\);
\STATE \hspace{0.5cm} \textbf{If} \(\tilde{G}^{(k + 1)} \leq \tilde{G}^{(k)}\)
\STATE \hspace{0.5cm} \hspace{0.5cm} Accept \(\mathbf{w}^{(k + 1)}\);
\STATE \hspace{0.5cm} \textbf{Else}
\STATE \hspace{0.5cm} \hspace{0.5cm} Accept \(\mathbf{w}^{(k + 1)}\) with probability \(p\) that depends on \(\mathbf{T}\);
\STATE \hspace{0.5cm} \textbf{End}
\STATE \textbf{Until Convergence}
\end{algorithmic}
\label{alg:ResNet_SA}
\end{algorithm}
It is worth noting that the composition of the training dataset plays a critical role in the prediction accuracy of the NN. If the dataset contains only weight vectors associated with relatively high radiation gains, the network may struggle to accurately predict the gains corresponding to low-gain (deep null) solutions. To mitigate this issue, a practical and effective approach is to collect training samples from the sequence of weight vectors generated by the SA algorithm applied to the theoretical radiation pattern. This sequence naturally spans a wide range of gain values, from high-gain initial solutions to low-gain solutions corresponding to deep nulls, thereby providing a balanced dataset for training.

\section{Simulation Results}
\begin{figure}[htp!]
\begin{center}
\includegraphics[scale=0.5]{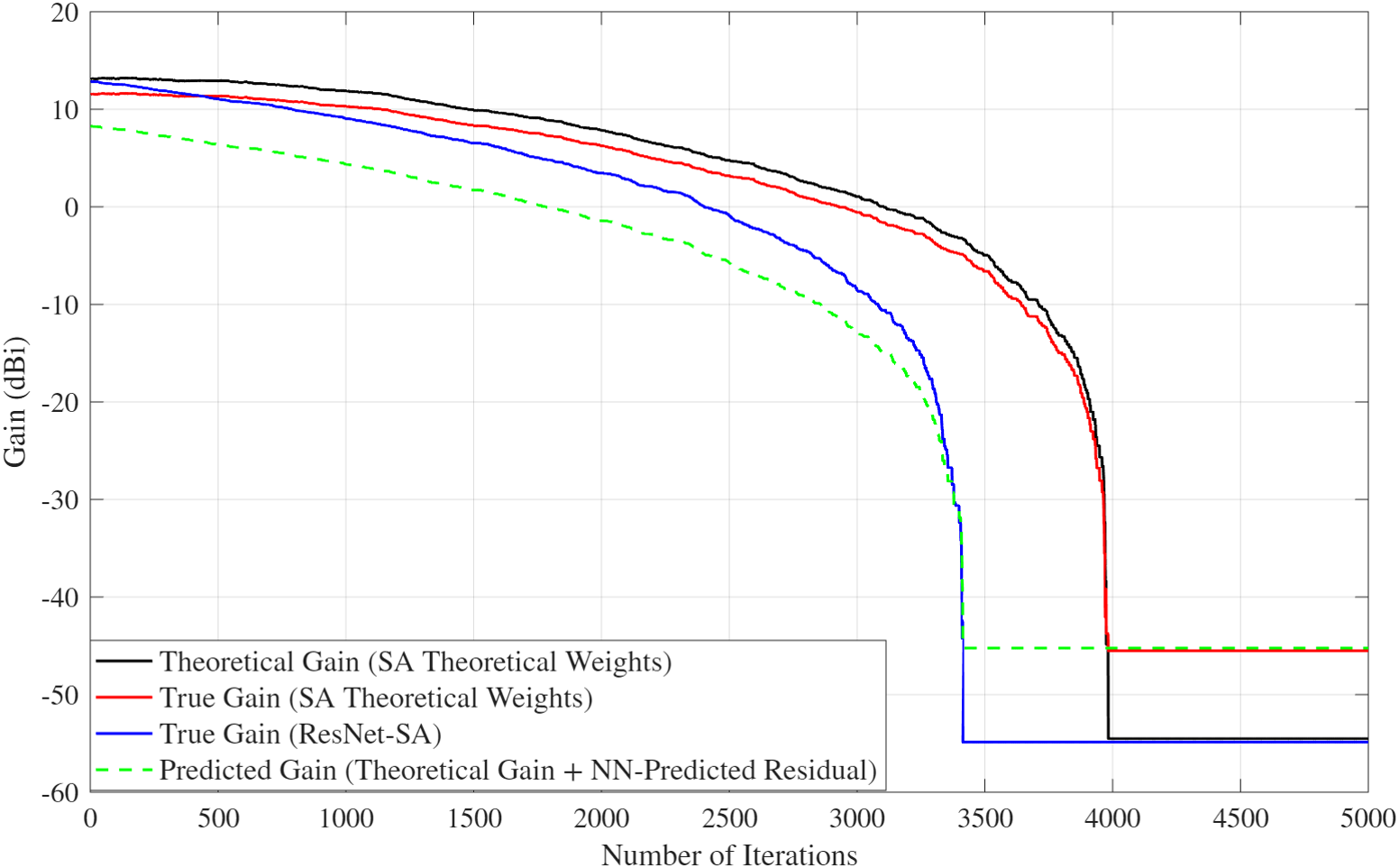}
\caption{Comparison of the achievable gain between the SA and the ResNet-SA algorithm. The angle to null is set to \(\psi = 2.5^{\circ}\) on the H-plane (\(\phi = 0^{\circ}\)). Number of examples in the training dataset is 4000. The black curve shows the convergence of the SA algorithm evaluated on the theoretical radiation pattern, while the red curve shows the corresponding gain of the true radiation pattern using weights optimized for the theoretical model. The blue curve illustrates the convergence of the ResNet-SA algorithm for the true radiation pattern, and the green curve represents the sum of the theoretical gain and the NN-predicted residual in the ResNet-SA algorithm. } 
\label{fig:True_gain_NN_25_1}
\end{center}
\end{figure}

\begin{figure}[htp!]
\begin{center}
\includegraphics[scale=0.5]{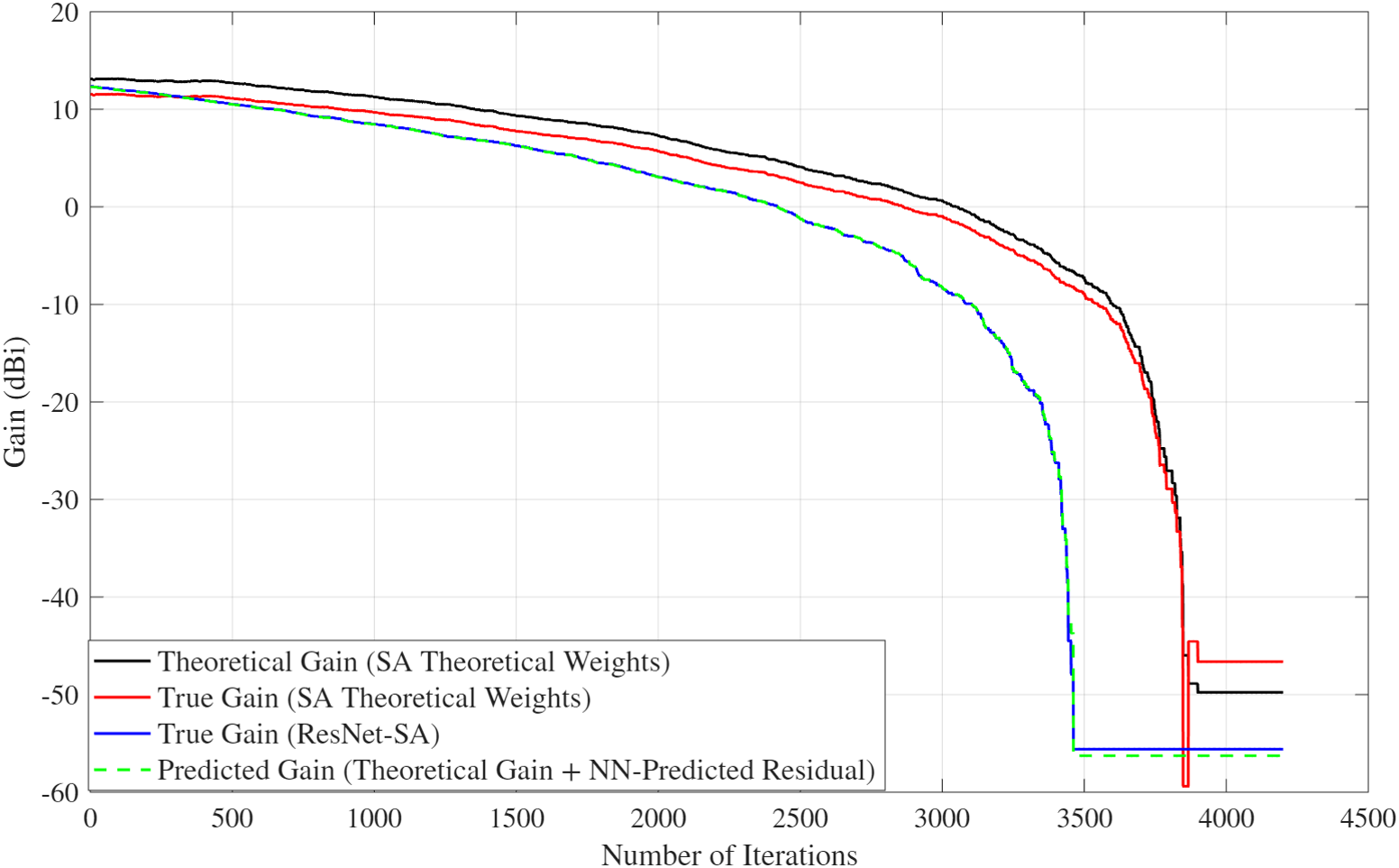}
\caption{Comparison of the achievable gain between the SA and the RL-SA algorithm. The angle to null is set to \(\psi = 2.5^{\circ}\) on the H-plane (\(\phi = 0^{\circ}\)). Number of examples in the training dataset is 80000. }
\label{fig:True_gain_NN_25_2}
\end{center}
\end{figure}

\begin{figure}[htp!]
\begin{center}
\includegraphics[scale=0.68]{pattern_25_NN.png}
\caption{Theoretical radiation pattern versus the true radiation pattern on H-plane \(\phi = 0^{\circ}\) with a null placed at \(\psi = 2.5^{\circ}\). The blue curve represents the true radiation pattern obtained using the weights optimized by the ResNet-SA algorithm. }
\label{fig:pattern_25_NN}
\end{center}
\end{figure}
To demonstrate the performance of the algorithms discussed in Section~\ref{sec:II} and to evaluate the achievable null depth for the interference direction, we present numerical results in this section. We consider an 18m diameter paraboloidal reflector operating at a frequency of 1.5GHz. The outer 0.5m rim of the reflector consists of 2756 contiguous reconfigurable segments. Each element is a unit cell of the reflectarray having side length \(0.5\lambda\). The feed model is given by (\ref{eq:mag_field}) with \(q = 1.14\) for the theoretical pattern and \(q = 1.5\) for the true pattern. The number of quantization levels of the weights is set to \(M = 4\).  \par
We first examine the prediction accuracy of the NN by applying the proposed ResNet-SA algorithm to optimize the RIS weights for the true radiation pattern at a single interference angle. The training dataset consists of sequences of weight vectors generated by the SA algorithm applied to the theoretical radiation pattern, spanning the optimization process from initialization to convergence. For each weight vector, both the corresponding theoretical gain and the measured true gain are included. In this example, the training data are collected for a single nulling angle \(\psi = 2.5^{\circ}\), and each sequence in the training set contains 4000 weight vectors. Fig.~\ref{fig:True_gain_NN_25_1} illustrates the convergence behavior of the theoretical gain, the true gain, and the neural-network-predicted gain when the training dataset consists of 4000 samples, corresponding to one full SA run for the theoretical electric field model. As observed, the null depth achieved by directly applying weights optimized for the theoretical pattern is degraded when evaluated using the true radiation pattern, as indicated by comparing the black and red curves. In contrast, the ResNet-SA algorithm converges to a  deeper null in the true pattern (blue curve), despite imperfect NN prediction of the gain mismatch (dashed green curve). This result reflects a scenario in which only a limited number of training samples are available. As the size of the training dataset increases, the NN’s accuracy in predicting the gain mismatch improves accordingly. This is demonstrated in Fig.~\ref{fig:True_gain_NN_25_2}, where the network is trained using 80000 samples. In this case, the predicted gain closely matches the true gain, enabling the ResNet-SA algorithm to more accurately guide the weight optimization process and achieve enhanced null depth in the true radiation pattern. Fig.~\ref{fig:pattern_25_NN} presents the H-plane co-pol pattern when the null is placed at \(\psi = 2.5^{\circ}\). As indicated by the previous two figures, the null depth is deeper for the weights generated by the RL-SA algorithm.

\section{Conclusion}
This report addressed RIS-assisted spatial nulling for large reflector antennas under the case when the true radiation pattern deviates from the theoretical model. Existing weight optimization algorithms, which assume accurate theoretical models, suffer from degraded null depth when applied to the true radiation pattern. To overcome this limitation, a residual learning network–assisted simulated annealing (ResNet-SA) framework was proposed, which leverages a NN to predict the gain mismatch between the theoretical and true antenna patterns without requiring explicit knowledge of the distorted electric field. Simulation results demonstrate that the proposed approach achieves deeper nulls in the true radiation pattern than directly applying weights optimized for the theoretical model, thereby validating the effectiveness of the ResNet-SA algorithm.

\bibliographystyle{IEEEtran}
\bibliography{mybibfile}

\end{document}